# Orbit-spin coupling and the circulation of the Martian atmosphere

James H. Shirley

*Jet Propulsion Laboratory, California Institute of Technology*

*Pasadena, CA, USA*

7 May 2016

Key points:

  We obtain an expression describing a weak coupling of orbital and rotational motions that yields accelerations within planetary atmospheres.

  Properties of the derived acceleration field are described and testable predictions are detailed.

  Orbit-spin coupling may contribute to the observed interannual variability of the circulation of the Martian atmosphere.

**Headings**

**Abstract**

**1. An unsolved problem of atmospheric physics**

**2. Orbital and rotational motions**

*2.1. Orbital motion*

*2.2. Rotational motion*

*2.3. Orbit-spin coupling?*

**3. Approach and derivation**

*3.1. Overview*

*3.2. Preliminaries: Rotation, revolution, and reference frames*

*3.3. Derivation*

*3.4 Substitution of variables*

*3.5. The coupling expression employing d**L**/dt*

*3.6. The coupling efficiency coefficient (c)*

**4. Properties of the coupling term and acceleration field**

*4.1. Partitioning between zonal and meridional components of acceleration*

*4.2. Temporal variability of the forcing function ($\dot{\boldsymbol{L}}$)*

*4.3. Acceleration magnitude and the coupling efficiency coefficient*

**5. What level of angular momentum exchange is permitted by solar system observations?**

**6. Predictability**

*6.1. Constructive and destructive interference effects*

*6.2. Range of variability*

*6.3. Intensification and relaxation of circulatory flows*

**Abstract:**

The physical origins of the observed interannual variability of weather and climate on Mars are poorly understood. In this paper we introduce a deterministic physical mechanism that may account for much of the variability of the circulation of the Mars atmosphere on seasonal and longer timescales. We derive a coupling expression linking orbital and rotational motions that produces an acceleration field varying with position and with time on and within a subject body. The spatially and temporally varying accelerations may interfere constructively or destructively with large-scale flows of geophysical fluids that are established and maintained by other means. The hypothesis predicts cycles of intensification and relaxation of circulatory flows of atmospheres on seasonal and longer timescales that are largely independent of solar forcing. The predictions of the hypothesis may be tested through numerical modeling. Examples from investigations of the atmospheric circulation of Mars are provided to illustrate qualitative features and quantitative aspects of the mechanism proposed. We briefly discuss the implications and applicability of the orbit-spin coupling hypothesis for planets other than Mars.

## 1. An unsolved problem of atmospheric physics

Spectacular global-scale dust storms (GDS) occur during the southern spring and summer seasons on Mars in some years but not in others [*Haberle*, 1986; *Kahn et al.*, 1992; *Zurek et al.*, 1992; *Zurek and Martin*, 1993; *Shirley*, 2015]. Multiple centers of dust-lifting appear to simultaneously become active in widely separated locations around the planet during the early stages of these events [*Martin and Zurek*, 1993; *Cantor et al.*, 2001; *Strausberg et al.*, 2005; *Wang and Richardson*, 2015]. Airborne dust typically encircles the globe within a few weeks, thereby profoundly affecting the temperature structure and the global circulation. The decay phases of such storms may last several months. Global-scale dust storms occurred on Mars in 2001 and in 2007, but the subsequent four Mars years (with perihelia in 2009, 2011, 2013, and late 2014) have been entirely free of such storms.

The origins of the interannual variability of global-scale dust storm occurrence on Mars are not well understood. As with the Earth, the circulation of the Mars atmosphere is driven by solar energy incident upon the system. This forcing is highly predictable and repeatable from year to year. In addition, the thermal time constant $\tau$ for the Mars atmosphere, which approximates the lifetime of a significant thermal perturbation introduced within the system, is quite short, at about 2 days [*Leovy*, 2001]. While this timescale provides scope for a quite remarkable range of thermal tidal phenomena [*Zurek*, 1976, 1981; *Zurek and Leovy*, 1981; *Wilson and Hamilton*, 1996; *Banfield et al.*, 2000; *Wilson*, 2002; *Wang et al.*, 2003; *Martinez-Alvarado et al.*, 2009; *Withers et al.*, 2011; *Kleinböhl et al.*, 2013; *Guzewich et al.*, 2014, 2016; *Wu et al.*, 2015], it appears to provide insufficient system memory to account for the observed persistence of anomalous atmospheric conditions (and some apparent teleconnections) over multi-day intervals (*e.g*., as evidenced by a nearly simultaneous initiation and continuation of dust lifting, from multiple centers, as observed early in the inception phases of GDS).

Mars atmospheric global circulation models (MGCMs) are remarkably effective in simulating the normal seasonal progression of atmospheric temperature and pressure on Mars [*Forget et al*., 1999; *Richardson et al*., 2007]. Such models attempt to directly incorporate or parameterize all of the relevant atmospheric physics. However, attempts to model the Mars dust cycle and GDS occurrence with MGCMs have met with only limited success [*cf. Pankine and Ingersoll*, 2002; *Basu et al*., 2006, and references therein; *Mulholland et al*., 2013; *Newman and Richardson*, 2015, and references therein). The intermittent occurrence of global-scale dust storms on Mars has accordingly been characterized as an outstanding unsolved problem of atmospheric physics [*Haberle*, 1986; *Kahn et al*., 1992*; Zurek et al*., 1992; *Zurek and Martin*, 1993; *Pankine and Ingersoll*, 2002, 2004; *Basu et al*., 2006; *Cantor*, 2007].

One possible explanation for our current inability to understand and model the GDS phenomenon on Mars is that our physical models may be in some way incomplete. In this paper we introduce and describe a physical mechanism that may represent a significant source of inter-annual variability for planetary atmospheres, and for the Mars atmosphere in particular.

Our approach to this problem is motivated and informed by recent empirical results documenting a remarkable systematic relationship linking the occurrence of past GDS on Mars with intervals in which the orbital angular momentum (*L*) of Mars (with respect to the solar system barycenter) is increasing or near peak values [*Shirley*, 2015]. A hypothetical coupling of the orbital and rotational angular momenta of Mars has been put forward as a possible explanation.

Below, in Section 3, we derive a mathematical expression describing a weak coupling between the orbital and rotational angular momenta of extended bodies in multi-body systems. Unlike all prior mechanisms for coupling extended body rotation states with orbital motions, including precession-nutation, and tidal friction, the process identified here does not in any way depend on tidal gradients of gravitational fields.

The derived orbit-spin coupling expression yields a global acceleration field that may interfere constructively or destructively with large-scale atmospheric motions originating due to other causes. (A closely related term, "spin-orbit coupling," is now widely employed in connection with studies of tidal friction. We therefore prefer to employ the term "orbit-spin coupling" to distinguish clearly between these two hypotheses). The predicted acceleration field pulsates in magnitude and periodically reverses direction. In the case of Mars, the mean cycle time of the putative dynamical forcing function is ~2.2 yr [*Shirley*, 2015], which is incommensurate with the 1.88 yr period of the Mars year that controls the progression of the seasons. A companion paper [*Shirley and Mischna*, 2016] examines the phase relationships of the solar irradiance and dynamical waveforms in detail. The complex phasing of the solar and dynamical contributions appears well suited to introduce inter-seasonal and inter-annual variability within the Mars atmospheric system [*Mischna and Shirley*, 2016; *Shirley and Mischna*, 2016].

The orbit-spin coupling hypothesis makes testable predictions, which are detailed here in Section 6. Hypothesis testing is best addressed through numerical modeling of the atmospheric circulation of a subject body. Efforts to test the predictions of the hypothesis through numerical modeling of the atmospheric circulation of Mars were initiated in 2014. The "coupling term acceleration" (as derived here) was incorporated within the dynamical core of the MarsWRF global circulation model [*Richardson et al.*, 2007]. The modified MGCM was thereafter subjected to extensive validation and testing. These experiments and their results are detailed in a companion paper [*Mischna and Shirley*, 2016].

The purposes of this paper are to document the method by which the orbit-spin coupling expression was obtained, and to describe the predictions of the associated orbit-spin coupling hypothesis. By so doing, it is anticipated that a wider involvement of the scientific community may be engaged, in order to further investigate the suspected coupling process and to better understand the physical interactions that are parameterized in an approximate way in the present version of the hypothesis. Our discussion (here and in companion papers) is sufficiently detailed to allow coding of algorithms and replication of the modeling work by other investigators using other GCMs.

In the following Section we briefly review salient aspects of the solar system dynamical environment within which the orbit-spin coupling process may operate. Our derivation is presented in Section 3, while in Section 4 we describe a number of key properties of the acceleration field. In Section 5 quantitative comparisons are employed to demonstrate that the operation of the proposed mechanism does not violate key constraints imposed either by astronomical observations or by high-precision ephemeris calculations. The discussions of Sections 4 and 5 provide a basis for the enumeration of specific predictions (Section 6) that may be validated (or invalidated) through further investigation. A discussion of key implications follows in Section 7, while our conclusions are detailed in Section 8. Brief reviews and discussions of salient findings from prior mainstream treatments of gravitation and rotation and related topics are provided in the Appendices, as this background material may be of interest to some readers. No new material is included within the Appendices.

## 2. Orbital and rotational motions

*2.1. Orbital motion*

As first recognized by Sir Isaac Newton [1687], the *orbital* motions of extended bodies that are members of two-body or multi-body systems are best understood with reference to the *center of mass* (or *barycenter*) of the subject system. Figure 1 illustrates orbital and rotational motions in a 2-body system. Here both bodies follow geometrically similar orbital trajectories about the center of mass (CM) of the pair, whose position lies along the line connecting the centers of the bodies. The dimensions of the two orbits may be obtained from the relations $M_1R_1 = M_2R_2$ and $D = R_1 + R_2$, where $M_1$, $M_2$ are the masses, $D$ is the distance separating the centers of the bodies, and $R_1$, $R_2$ are the distances from the centers of the bodies to their common barycenter [*Roy*, 1978].

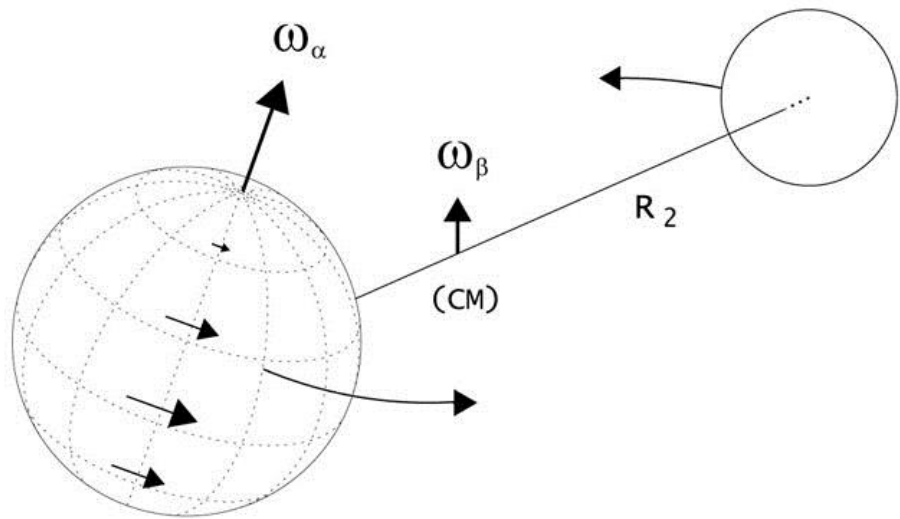

**Figure 1.** Rotation and revolution in a 2-body system. The curved arrows represent the orbital trajectories of a subject body (at left) and its companion as they revolve about the center of mass (CM) or *barycenter* of the pair. The orbital motion is also represented by the angular velocity vector $\omega_\beta$, which is normal to the orbital plane. **R** denotes the orbital radius extending from the body center to the barycenter, here labeled only for the companion body. The axial rotation of the subject body is parameterized by the angular velocity vector $\omega_\alpha$. Barycentric orbital motions of distant stars are occasionally detected from Earth; such observations are routinely employed in the detection and characterization of extrasolar planets [*e.g.*, *Marcy and Butler*, 1998].

A situation similar to that depicted in Fig. 1 obtains for multi-body systems such as our solar system. The Sun and the planets perform complex orbits about the *solar system barycenter*, as the directions of the planets in space vary with time [*Newton*, 1687; *Jose*, 1965; *Fairbridge and Shirley*, 1987; *Charvátová and Střeštík*, 1991]. The solar system barycenter by definition represents the origin of the *solar system inertial frame*; this is the most appropriate coordinate system for dynamical calculations of planetary motions (Newton's laws of motion are only valid in inertial frames). The solar system inertial frame is the fundamental underlying coordinate system assumed throughout the following discussion.

In the simplified example of Fig. 1, the barycentric orbit plane and the equatorial plane of the subject body are not parallel. The angle of inclination of these two planes is termed the *obliquity* ($\varepsilon$). Another key aspect of the orbital motions of the bodies illustrated is characterized by the orbital angular velocity vector $\omega_\beta$, which is by definition normal to the orbital plane; this quantity will play an important role in the derivation of Section 3 below.

The barycentric angular momentum (***L***) of the orbital motion, as considered later here, is most simply obtained through the equation ***L*** = M***VR***, where M is the mass of the subject body, ***V*** is the orbital velocity, and ***R*** is the distance separating the center of the subject body from the system barycenter. While the total angular momentum of the solar system is conserved, that of the individual members of the system is variable. Mars thus gains orbital angular momentum during some intervals, while during others it is yielding up orbital angular momentum to other members of the solar system family. We suspect that the Mars atmosphere may participate, either actively or passively, in this exchange [*Shirley*, 2015].

Relationships linking the variability of the orbital angular momentum of Mars with the occurrence of GDS on Mars were previously described in *Shirley* [2015], where a more complete description of the origins of the variability of planetary orbital angular momenta in the solar system may be found.

*2.2. Rotational motion*

The *rotational* motion of the subject body (depicted at left in Fig. 1) is indicated by the bold arrows superimposed on the surface of the subject body, and by the axial rotation angular velocity vector $\omega_\alpha$. As indicated, the rotational velocity is larger at locations nearer the equator than in higher latitudes. This illustrates a fundamental qualitative difference between orbital and rotational motions that plays a role of some importance. In rotation, the constituent particles of an extended body move in *concentric* trajectories with velocities that are dependent on the distance of the particle from the axis of rotation (Fig. 1). However, in orbital revolution, the constituent particles of the subject body are considered to move in *parallel* trajectories, with identical velocities [*Darwin*, 1898; *Shirley*, 2006]. It is thus problematic to employ purely rotational concepts and relations when considering orbital motions. A more detailed discussion of this topic may be found in *Shirley* [2006].

*2.3. Orbit-spin coupling?*

Results cited earlier [*Shirley*, 2006, and references therein; *Shirley*, 2015] raise questions of the possible existence of an unknown mechanism capable of in some way coupling the orbital and rotational motions of extended bodies, as depicted in Fig. 1. How might such a mechanism operate, and how may we further explore this possibility? Existing theory provides little encouragement here. For most dynamical calculations, the orbital and rotational motions of extended bodies are considered to be entirely independent and uncoupled (Appendix 1). Over long time periods, the rotation states of some extended bodies may be modified significantly through the action of tidal forces (Appendix 2). However, the presently known tidal coupling mechanisms appear to be of little relevance to our problem. In particular, the phenomena of precession-nutation, and tidal friction, are not typically considered to play any significant role in the determination of atmospheric motions. In addition, the gravitational tides on Mars are both qualitatively and quantitatively inadequate to account for the relationships previously found linking GDS occurrence with variations in Mars' orbital angular momentum (Appendix 3).

An important constraint from observations significantly limits the solution space for the orbit-spin coupling problem. The *orbital* angular momenta of planets in the solar system are typically far larger than the corresponding *rotational* angular momenta. In the case of Mars, the orbital angular momentum has a mean value of ~ 3.51 x $10^{39}$ kg m$^2$ s$^{-1}$, while the rotational angular momentum is more than 8 orders of magnitude smaller, at ~1.91 x $10^{31}$ kg m$^2$ s$^{-1}$. From this comparison, it is apparent that any possible coupling between these reservoirs must necessarily be quite weak; if this were not the case, then the rotational motions of Mars (and the other planets) would be strongly perturbed in ways that are not observed. We will return to this question in Section 5, where we estimate the levels of angular momentum exchange that may be allowed by observations.

## 3. Approach and derivation

### 3.1. Overview

Our goal is to introduce the *rotary component of orbital motion* in a dynamical model that also incorporates the phenomena of rotation. To do this we will employ as a template the derivation of inertial forces introduced by *Coriolis* [1835]. A minor modification to the classical approach leads to an expression that includes the cross product of the angular velocities of orbital revolution ($\omega_\beta$) and axial rotation ($\omega_\alpha$). The expression thereby obtained cannot accurately be characterized as a "new" inertial force in the traditional sense, due in part to the qualitative dissimilarity of orbital and rotational motions, as noted in passing above in Section 2.2. The expression obtained is nonetheless quite useful as a stepping-stone; with subsequent further modifications, in Sections 3.4-3.6, we obtain a quantitative expression describing a weak coupling of the orbital and rotational motions that is suitable for evaluation through numerical modeling.

### 3.2. Preliminaries: Rotation, revolution, and reference frames

In the following derivation, we mainly follow the nomenclature and approach employed previously in derivations of inertial forces by *Goldstein* [1959] and *Marion and Thornton* [1988], who obtain:

$$\mathbf{F}_{\text{effective}} = m\ d/dt\ \mathbf{v} = \mathbf{F}_{\text{Total}} - m\ \boldsymbol{\omega} \times (\boldsymbol{\omega} \times \mathbf{r}) - 2\ m\ \boldsymbol{\omega} \times \mathbf{v} - m\ (d/dt\ \boldsymbol{\omega}) \times \mathbf{r} \qquad (1)$$

That is, the effective force on a particle in a reference frame rotating with angular velocity $\boldsymbol{\omega}$ consists of the total of the applied forces, together with three inertial terms. These are respectively the centrifugal force, the Coriolis force, and a comparatively insignificant (and often neglected) force arising due to angular accelerations.

The position vector $\mathbf{r}$ of equation (1) is identified in Fig. 2. The figure shows a standard body-fixed Cartesian coordinate system with origin at the center of a subject body. We will refer to this system as the "$\alpha$ frame." The x- and y-axes lie within the equatorial plane, and the z-axis coincides with the axis of rotation. The position vector $\mathbf{r}$ points to a particular location (P) on or within the body. The point P may be in motion with respect to the $\alpha$ frame; for instance, we might wish to identify P with a projectile in flight. We denote that velocity with the letter $\mathbf{v}$ (as it appears in equation 1), or equivalently with the notation $d/dt\ \mathbf{r}$. The $\alpha$ frame rotates with respect to inertial frames with the angular velocity of the axial rotation $\boldsymbol{\omega}_\alpha$.

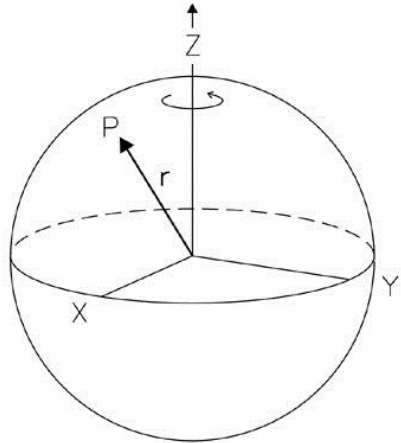

**Figure 2.** The α frame. The equator of the subject body provides the fundamental (x-y) plane for a standard body-fixed Cartesian coordinate system that rotates with the angular velocity $\omega_\alpha$. The position vector **r** identifies a particular location P within this system.

We now define a second Cartesian coordinate system, this time with origin at the barycenter of the 2-body system of Fig. 1. Suppose that the line connecting the centers of the two bodies defines the x-axis of this system, and that the z-axis coincides with the barycentric angular velocity vector, $\omega_\beta$. Like the α frame, this system rotates with respect to the stellar background. We will refer to this system as the "β frame."

Consider the viewpoint of an observer who is stationary with respect to the β frame. Suppose this "β observer" is viewing the subject body from a position somewhere along the line connecting the centers of the two bodies (refer once again to Fig. 1). This individual will observe the orbital motion of the subject body with respect to the stellar background, and will, in addition, see particular features on the surface of the subject body rotating into view at a rate that differs from the rotation rates obtained by inertial or α frame observers. This relative angular velocity $\omega_\rho$ is the difference of the axial and barycentric rates, i.e.,

$$\omega_\rho = \omega_\alpha - \omega_\beta \tag{2}$$

It is important to note here that the angular rate, $\omega_\rho$, determined by the β observer has no dependence on the distance of the body or the observer from the barycenter. The β frame velocity of a point P on the surface of the subject body is obtained most readily from the relation $\omega_\rho \times \mathbf{r}$, where $\mathbf{r}$ continues to represent the body-centered position vector of Fig. 2. The relative rotation ($\omega_\rho$) defined here is closely related to the synodic rotation rate parameter of solar astronomy, which is simply the rate of rotation of the Sun as viewed from the moving platform of the Earth.

Our β observer concludes that the total rotary motion of a point P on the subject body may be represented as the sum of the two rotary velocities $\omega_\rho \times \mathbf{r}$ and $\omega_\beta \times \mathbf{r}$. Although the resulting total rotational velocity with respect to inertial frames is numerically and directionally equivalent to the conventional description, where the rotation velocity is given by $\omega_\alpha \times \mathbf{r}$, the information content of these two descriptions is not the same: The β observer's description includes quantitative information on the angular misalignment and the relative magnitudes of the rotary motions of rotation and revolution.

Before moving forward we must take note of one other key feature of the physical system of Fig. 1. Suppose first that the subject body axial rotation rate, $\omega_\alpha$, does not vary with time. If so then (2) dictates that changes in the angular velocity $\omega_\beta$ must be accompanied by changes in $\omega_\rho$ that are identical in magnitude and opposite in sign (i.e., $d/dt\ \omega_\beta = -d/dt\ \omega_\rho$). Next suppose that the axial rotation of the subject body *does* vary with time, due to unrelated causes (i.e., $d/dt\ \omega_\alpha \neq 0$). In this more general case, the accelerations of $\omega_\alpha$ will be detected by the $\beta$ observer as a component of the time-variability of $\omega_\rho$ that is independent of any changes due to time-variability of $\omega_\beta$ ; that is,

$$d/dt\ \omega_{\rho\ observed}\ =\ d/dt\ \omega_{\rho\ orbital}\ +\ d/dt\ \omega_\alpha\ . \qquad (3)$$

We will keep track of both sources of acceleration in the following derivation.

As we noted in connection with equation (2), the orbital radius vector ($\mathbf{R}_1$) is not required in the description of the rotary motions of the subject body with respect to the $\beta$ frame observer or with respect to inertial frames. It plays no active role in the *Goldstein* [1959] or *Marion and Thornton* [1988] derivations of inertial forces, and is accordingly neglected by most authors. In the following derivation we will follow that tradition and (temporarily) collapse this to zero length, in effect shifting the point of origin of the $\omega_\beta$ vector of Fig. 1 to coincide with the center of the subject body. Figure 3 depicts the resulting relationship of the axial, barycentric, and relative angular velocity vectors, as given earlier in equation (2). While it is convenient to neglect $\mathbf{R}_1$ in the present context, we will be required to re-introduce this factor when our focus shifts to considerations of angular momenta.

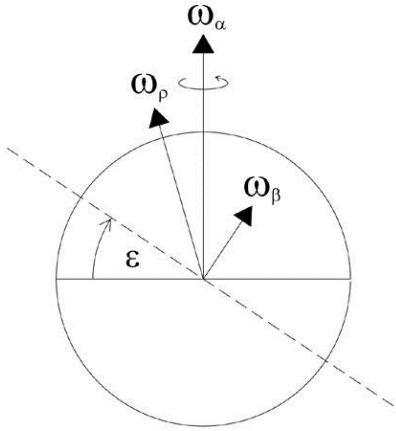

**Figure 3.** Angular relationships. In this view the 3 angular velocity vectors lie within the plane of the figure. The horizontal line represents the equator of the subject body, with the axial rotation vector $\omega_\alpha$ directed toward the top of the figure. The dashed line is the barycentric orbit plane, and $\omega_\beta$ is the angular velocity of the barycentric revolution. The inclination of these planes is the obliquity $\varepsilon$. The relative rotation vector $\omega_\rho$ is obtained by subtraction (i.e., $\omega_\rho = \omega_\alpha - \omega_\beta$; equation (2)).

*3.3. Derivation*

From the β observer's point of view, we appear to have all the necessary ingredients for producing something like a Coriolis force: We have a rotating reference frame tied to the motion of a physical system, and we have test particles (or, mass elements) moving with significant velocities with respect to that rotating frame.

The following derivation differs from those of *Goldstein* [1959] and *Marion and Thornton* [1988] in the following way. In the usual procedure we first resolve velocities and accelerations in the inertial frame, and then we transform (or "spin up") the coordinate system to obtain the corresponding accelerations in the α frame. In the present derivation we obtain the accelerations in two steps. We begin with velocities and accelerations in an inertial frame as before, but we next resolve the relevant accelerations *in the β frame*. The final solution is then obtained by a second transformation, from the β frame to the α frame.

In the textbook derivations, one begins by specifying all of the velocities of the subject body with respect an inertial frame, i.e.,

$$\mathbf{V} = \mathbf{v}_t + \boldsymbol{\omega}_\alpha \times \mathbf{r} + \mathbf{v}_\alpha. \tag{4}$$

where $\mathbf{V}$ denotes the total velocity, $\mathbf{v}_t$ represents the sum of the translational velocities for the center of the subject body, $\boldsymbol{\omega}_\alpha \times \mathbf{r}$ describes the rotation of the chosen noninertial frame, and $\mathbf{v}_\alpha$ is a velocity measured with respect to that rotating (α) frame. The $\mathbf{v}_t$ are purely translational motions; as they are traditionally considered to have no rotary component, these play no role in the subsequent development. We are thus free to discard the first term. The initial velocity specification for the derivation is then [*Goldstein*, 1959, equation 4-104]:

$$\mathbf{V} = \boldsymbol{\omega}_\alpha \times \mathbf{r} + \mathbf{v}_\alpha \tag{5}$$

The first step of the present derivation is to express the above initial velocity in an analogous and numerically equivalent form employing the β frame:

$$\mathbf{V} = \boldsymbol{\omega}_\beta \times \mathbf{r} + \mathbf{v}_\beta \tag{6}$$

The numerical equivalence of (5) and (6) allows us to immediately solve for $\mathbf{v}_\beta$, the velocity relative to the β frame:

$$\mathbf{v}_\beta = \omega_\alpha \times \mathbf{r} - \omega_\beta \times \mathbf{r} + \mathbf{v}_\alpha = \omega_\rho \times \mathbf{r} + \mathbf{v}_\alpha, \qquad (7)$$

where we recognize the β observer's relative rotation velocity $\omega_\rho \times \mathbf{r}$. Our initial inertial frame velocity is then explicitly

$$\mathbf{V} = \omega_\beta \times \mathbf{r} + \omega_\rho \times \mathbf{r} + \mathbf{v}_\alpha \qquad (8)$$

We next obtain the time derivative of (8):

$$d/dt\, \mathbf{V} = \omega_\beta \times d/dt\, \mathbf{r} + (d/dt\, \omega_\beta) \times \mathbf{r} + \omega_\rho \times d/dt\, \mathbf{r} + (d/dt\, \omega_\rho) \times \mathbf{r}$$

$$+ d/dt\, \mathbf{v}_\alpha \qquad (9)$$

As noted in equation (3) the time derivative of the relative rotation $\omega_\rho$ may include contributions from two sources (i.e., $d/dt\, \omega_{\rho\,\text{observed}} = d/dt\, \omega_{\rho\,\text{orbital}} + d/dt\, \omega_\alpha$). We replace the fourth term on the right side of (9) with (3). This substitution allows us to drop the "observed" and "orbital" subscripts, as we are now accounting for the two sources separately. In all that follows, $\omega_\rho$ will refer only to the orbitally determined value of equation (2).

As in the textbook derivations, we next make use of the following general equation that relates the time derivative of an arbitrary vector, *q*, in some specific starting frame to the corresponding vector quantity in a system that is rotating with angular velocity $\boldsymbol{\omega}$ relative to the original frame:

$$d/dt \; \boldsymbol{q}_{\text{original frame}} \; = d/dt \; \boldsymbol{q}_{\text{rotating frame}} + \boldsymbol{\omega} \times \boldsymbol{q} \; . \tag{10}$$

To resolve the accelerations in the β frame, we apply (10) to all of the terms on the right side of the fully expanded version of (9). Term by term, we obtain:

$$\boldsymbol{\omega}_\beta \times d/dt \; \mathbf{r} \; = \; \boldsymbol{\omega}_\beta \times (\boldsymbol{\omega}_\beta \times \mathbf{r}) + \boldsymbol{\omega}_\beta \times \mathbf{v}_\beta$$

$$= \; \boldsymbol{\omega}_\beta \times (\boldsymbol{\omega}_\beta \times \mathbf{r}) + \boldsymbol{\omega}_\beta \times (\boldsymbol{\omega}_\rho \times \mathbf{r}) + \boldsymbol{\omega}_\beta \times \mathbf{v}_\alpha, \tag{11a}$$

$$(d/dt \; \boldsymbol{\omega}_\beta) \times \mathbf{r} \; = \; (d/dt \; \boldsymbol{\omega}_\beta) \times \mathbf{r} + (\boldsymbol{\omega}_\beta \times \boldsymbol{\omega}_\beta) \times \mathbf{r} = (d/dt \; \boldsymbol{\omega}_\beta) \times \mathbf{r}, \tag{11b}$$

$$\boldsymbol{\omega}_\rho \times d/dt \; \mathbf{r} \; = \; \boldsymbol{\omega}_\rho \times (\boldsymbol{\omega}_\beta \times \mathbf{r}) + \boldsymbol{\omega}_\rho \times \mathbf{v}_\beta$$

$$= \; \boldsymbol{\omega}_\rho \times (\boldsymbol{\omega}_\beta \times \mathbf{r}) + \boldsymbol{\omega}_\rho \times (\boldsymbol{\omega}_\rho \times \mathbf{r}) + \boldsymbol{\omega}_\rho \times \mathbf{v}_\alpha, \tag{11c}$$

$$(d/dt \; \boldsymbol{\omega}_\rho) \times \mathbf{r} \; = \; (d/dt \; \boldsymbol{\omega}_\rho) \times \mathbf{r} + (\boldsymbol{\omega}_\beta \times \boldsymbol{\omega}_\rho) \times \mathbf{r}, \tag{11d}$$

$$(d/dt \; \boldsymbol{\omega}_\alpha) \times \mathbf{r} \; = \; (d/dt \; \boldsymbol{\omega}_\alpha) \times \mathbf{r} + (\boldsymbol{\omega}_\beta \times \boldsymbol{\omega}_\alpha) \times \mathbf{r}, \tag{11e}$$

$$d/dt \; \mathbf{v}_\alpha \; = \; d/dt \; \mathbf{v}_\alpha + \boldsymbol{\omega}_\beta \times \mathbf{v}_\alpha \tag{11f}$$

where $d/dt \; \mathbf{r}$ is recognized as $\mathbf{v}_\beta$, and where we have fully expanded all instances of $\mathbf{v}_\beta$ using equation (7). Collecting and rearranging terms, we have the following β frame accelerations:

$$d/dt \; \mathbf{V} = d/dt \; \mathbf{v}_\alpha \; + \; (d/dt \; \boldsymbol{\omega}_\beta) \times \mathbf{r} \; + (d/dt \; \boldsymbol{\omega}_\rho) \times \mathbf{r} \; + (d/dt \; \boldsymbol{\omega}_\alpha) \times \mathbf{r}$$

$$+ \; \boldsymbol{\omega}_\beta \times (\boldsymbol{\omega}_\beta \times \mathbf{r}) + \; \boldsymbol{\omega}_\rho \times (\boldsymbol{\omega}_\rho \times \mathbf{r})$$

$$+ \ 2 \ \omega_\beta \times v_\alpha \ + \ \omega_\rho \times v_\alpha$$

$$+ \ \omega_\beta \times (\omega_\rho \times r) \ + \ \omega_\rho \times (\omega_\beta \times r)$$

$$+ \ (\omega_\beta \times \omega_\rho) \times r \ + \ (\omega_\beta \times \omega_\alpha) \times r \tag{12}$$

The two angular acceleration terms on the first line of (12) involving $\omega_\beta$ and $\omega_\rho$ yield linear accelerations that are opposite and equal (as before, $d/dt \ \omega_\beta = - \ d/dt \ \omega_\rho$). We simplify the equation by canceling these terms.

In order to resolve the accelerations in the $\alpha$ frame we must make use once more of (10). In this case the appropriate spin-up angular velocity is $\omega_\rho$. For the remaining time derivatives on the first line of (12) we write

$$d/dt \ v_{\alpha \ (\beta \ frame)} \ = \ d/dt \ v_{\alpha \ (\alpha \ frame)} \ + \ \omega_\rho \times v_\alpha \tag{13a}$$

$$(d/dt \ \omega_\alpha) \times r \ = \ (d/dt \ \omega_\alpha) \times r \ + \ (\omega_\rho \times \omega_\alpha) \times r \tag{13b}$$

The final term of (13b) is the negative of the final term of (12), allowing us to cancel these. The collected $\alpha$ frame accelerations are:

$$d/dt \ V \ = \ d/dt \ v_\alpha \ + \ (d/dt \ \omega_\alpha) \times r \ + \ \omega_\rho \times (\omega_\rho \times r) \ + \ \omega_\beta \times (\omega_\beta \times r)$$

$$+ \ \omega_\beta \times (\omega_\rho \times r) \ + \ \omega_\rho \times (\omega_\beta \times r) \ + \ 2 \ \omega_\beta \times v_\alpha \ + 2 \ \omega_\rho \times v_\alpha$$

$$+ \ (\omega_\beta \times \omega_\rho) \times r \tag{14}$$

The $\omega_\beta$ and $\omega_\rho$ terms of equation (14) involving $v_\alpha$ may be combined to yield the usual $\alpha$ frame Coriolis acceleration:

$$2\,\omega_\beta \times \mathbf{v}_\alpha + 2\,\omega_\rho \times \mathbf{v}_\alpha = 2\,\omega_\alpha \times \mathbf{v}_\alpha. \tag{15a}$$

And, making use of equation (2) in the form $\omega_\alpha = \omega_\rho + \omega_\beta$, we can further combine and simplify four of the vector triple products of (14):

$$\omega_\rho \times (\omega_\rho \times \mathbf{r}) + \omega_\beta \times (\omega_\beta \times \mathbf{r}) + \omega_\beta \times (\omega_\rho \times \mathbf{r}) + \omega_\rho \times (\omega_\beta \times \mathbf{r}) =$$

$$\omega_\alpha \times (\omega_\alpha \times \mathbf{r}) \tag{15b}$$

This is the $\alpha$ frame centrifugal acceleration of the standard solution.

Continuing to follow the textbook derivations, we may now express everything in terms of forces, making use of $\mathbf{F} = m\mathbf{a}$ and rearranging terms to isolate the effective forces. Introducing the mass $m$ on both sides, $d/dt\,\mathbf{V}$ is identified with the sum of all forces acting ($\mathbf{F}_{Total}$); we write

$$\mathbf{F}_{effective} = m\,d/dt\,\mathbf{v}_\alpha = \mathbf{F}_{Total} - m\,\omega_\alpha \times (\omega_\alpha \times \mathbf{r}) - 2\,m\,\omega_\alpha \times \mathbf{v}_\alpha$$

$$- m\,(d/dt\,\omega_\alpha) \times \mathbf{r} - m\,(\omega_\beta \times \omega_\rho) \times \mathbf{r} \tag{16}$$

The first four terms on the right side of (16) are identical to those of the standard solution for inertial forces of equation (1). These need not be discussed further. The final term is entirely new.

Anticipating future needs, we can usefully re-formulate the angular velocities cross product of this new term by making use of equation (2) together with the distributive law. If $\omega_\rho$ is represented in the equivalent form ($\omega_\alpha - \omega_\beta$), we immediately recognize that the cross product of $\omega_\beta$ with itself yields a null vector. Thus we can more simply represent the cross product of the new term in the equivalent form $\omega_\beta \times \omega_\alpha$. In addition, for brevity, we will henceforth drop the mass variable *m*, and speak exclusively of accelerations. The newly obtained acceleration becomes:

$$-(\omega_\beta \times \omega_\alpha) \times \mathbf{r} \qquad (17)$$

Equation (17) is of considerable interest for the coupling problem in general, as the cross product (within parentheses) explicitly involves the angular velocities of orbital revolution and axial rotation. But what is the nature of the indicated acceleration? Mathematically, the cross product of the angular velocities yields a vector that is orthogonal to both. This vector must thereby lie within the equatorial plane of the subject body. Crossing this vector with a position vector (**r**) then produces a new vector that is tangential to the surface of the body at the location specified by **r**.

Figure 4 displays a typical solution for relative magnitudes and directions of the acceleration given by (17) over the surface of a subject body. We must visualize the axial rotation of the body carrying a particular surface location through this pattern (or *field*) of accelerations once per day. Consider the vectors plotted for points on the northernmost latitude circle (i.e., 60° N). As the body rotates, the acceleration points northward, then eastward, then southward, then westward, varying systematically in magnitude through this interval. Accelerations cycling in this way may potentially give rise to circulatory motions of geophysical and astrophysical fluids. For convenience, in later Sections and in companion papers, we will occasionally refer to this rotary progression of the direction of acceleration as the "quasi-diurnal component" of variability.

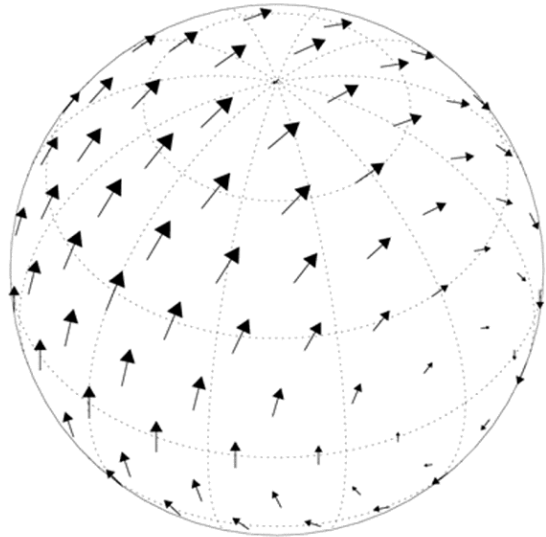

**Figure 4.** Vector representation of the accelerations imparted by (17) over the surface of an extended body. The lengths of the displayed vectors are proportional to their magnitude. Latitude and longitude grid lines at 30° intervals are shown for reference. The equally spaced grid points are obtained from an icosahedral system. (The acceleration vector for the north polar location has been omitted).

The accelerations disappear at locations where **r** is parallel to the vector product $\omega_\beta \times \omega_\alpha$. One such location is seen at the lower right of the figure. Peak accelerations are attained on the great circle of longitude that is 90° removed from this location. The accelerations are directed exclusively northward (or southward) everywhere along this great circle. This aspect is strongly reminiscent of the force diagram for a classical mechanical couple, such as a belt and pulley system, in which oppositely-directed accelerations applied at some distance from each other provide a turning tendency about an axis. We are thus fully justified in identifying (17) as a "coupling term."

From this point forward it will be convenient to label the pattern of accelerations illustrated in Fig. 4 as an "acceleration field." We will also henceforth consider the values of the position vector **r** to represent a set of fixed grid points, rather than to represent some arbitrary location corresponding to a particle that may be in motion, as previously described in Section 3.2 and Fig. 2.

We are still several steps removed from the desired final expression for coupling the orbital angular momentum with the rotational motion, as called for in *Shirley* [2006] and *Shirley* [2015]. In particular, for physical problems involving orbital velocities, orbital energy, or angular momenta, it is necessary to re-introduce factors of the orbital radius vector **R**. The representation of orbital motion in (17) employing the parameter $\omega_\beta$ in isolation lacks this factor and thus provides an incomplete description of the physical reality we wish to represent.

*3.4 Substitution of variables*

The orbital angular momentum (**L** = M**VR**) may also be written in the equivalent form **L** = M ($\omega_\beta \times$ **R**) × **R**. The angular momentum and angular velocity vectors are parallel by definition, and thus the additional factors essentially serve to scale the magnitude. Thus we may with confidence substitute **L** for $\omega_\beta$ in equation (17), thereby achieving our nominal objective of re-introducing some factor of the orbital radius vector within our still incomplete solution for orbit-spin coupling:

$$-(\boldsymbol{L} \times \omega_\alpha) \times \mathbf{r} \qquad (18)$$

(For some applications, it may also be necessary to replace the rotational angular velocity $\omega_\alpha$ in (18) with the rotational angular momentum, given by $I\omega_\alpha$, where $I$ is the inertia tensor of the planet in question. This additional approximately constant factor is not required for our purposes. Additionally, hereafter and in companion papers, we assign a value of unity to the Mars mass for numerical representation and calculations involving $L$).

The above coupling expression is still suboptimal from the standpoint of numerical modeling. We are principally interested in the *variability with time* of the orbital angular momentum [*Shirley*, 2015], as a forcing function, rather than in its gross value, which is numerically quite large. To resolve the temporal variability using (18) will require us to form small differences of large (vector) quantities, for all grid points, and for all time steps considered. We may avoid these extra operations by instead employing the rate of change of the orbital angular momentum (*dL/dt*) directly.

*3.5. The coupling expression employing dL/dt*

The rotational angular velocity $\omega_\alpha$ and the gridded position vectors **r** of (18) are effectively invariant over short periods of time. The time derivative of (18) is therefore quite easily obtained. After taking the derivative, the coupling term may be rewritten in the following form, where we have for brevity replaced the variable *dL/dt* with an alternate notation (i.e., $\dot{L}$):

$$- (\dot{L} \times \omega_\alpha) \times \mathbf{r} \tag{19}$$

We have introduced an extra factor of inverse time with this step (the dimensions of (19) are now kg m$^2$ s$^{-3}$). Expression (19) thus represents the *rate of change of the torque of the couple*. (This is analogous to the rate of change of force in ordinary mechanical problems). For numerical modeling, we wish to specify an acceleration value for each grid point and time step. It is a simple matter to obtain the local acceleration (at each grid point) from its time derivative. For convenience, in practice, we employ the numerical value of $\dot{L}$ obtained directly from ephemeris calculations [*Mischna and Shirley*, 2016] in (19). An integration over *t*=0 to 1 sec is implied but not explicitly performed. As the $\dot{L}$ waveform represents the principal source of variability for the coupling term, we will occasionally refer to this as the "forcing function" for the couple.

The issue of the very large magnitude of the accelerations (Section 2.3) remains to be addressed.

*3.6. The coupling efficiency coefficient (c)*

Coupling of the orbital and rotational angular momenta of solar system bodies is constrained by observations to be quite weak and inefficient (*cf.* Section 2.3). To address this we next introduce a scalar "coupling efficiency coefficient" (*c*). Our final expression for the coupling term then becomes

$$- c\, (\dot{L} \times \boldsymbol{\omega}_a) \times \mathbf{r} \qquad (20)$$

The parameter *c* quantifies the fractional value of the "orbital torque" applied to the totality of all of the concentric rotating "shells" of the subject body. Some portion of this may be available for accelerating the atmosphere under the present hypothesis. *c* plays a role here that is in some ways similar to that of the coefficient of friction $\mu$ in mechanics problems. $c$, like $\mu$, operates on and represents a fractional proportion of a dynamical quantity. $c$ is likely to take on different values depending on the physical properties and nature of the systems under investigation. In common with $\mu$, it may be viewed as a placeholder for representing multiple physical interactions operating on various scales, some of which may be dissipative. We accordingly consider $c$ to represent an intrinsic physical property of a particular subject body, whose value may be constrained through observations and modelling (*Mischna and Shirley*, 2016).

### 4. Properties of the coupling term and acceleration field

The coupling term (20) is a vector triple product involving orbital and rotational vector quantities and the position vector **r**. As we have previously noted, because it is a cross product with **r**, it acts only in directions that are *perpendicular* to the radius vector of the body (see Fig. 4). This is of some importance, as it dictates that no acceleration due to the couple may act along the radius vector, approximately parallel to the gravitational acceleration ***g***. This property is consistent with and required by observations. A second fundamental property of interest is that the accelerations of (20) must disappear completely if the $\dot{\boldsymbol{L}}$ and $\boldsymbol{\omega_\alpha}$ vectors are aligned (as their cross product disappears when they are parallel or antiparallel in orientation).

*4.1. Partitioning between zonal and meridional components of acceleration*

The meridional (north and south) components of the acceleration field of Fig. 4 appear by inspection to be larger than the zonal (east and west) components over most of the surface. We can quantify this potentially important spatial tendency, both globally and locally, through summation and averaging of the meridional and zonal components as illustrated in Fig. 4. The grid employed for Fig. 4 consists of 162 equidistant locations distributed over the sphere. These are arranged along a set of small circles of latitude, at ±75°, ±60°, ±45°, ±30°, and ±15°, together with the equator, and at each pole. Summation of the absolute values of the meridional and zonal components for all grid points reveals that the instantaneous global ratio of zonal (east-west) accelerations to meridional (north-south) accelerations is ~0.5, confirming the suspected dominance of meridional components over the zonal components globally.

We next examine the relative magnitudes of the zonal and meridional components as a function of latitude, as illustrated in Fig. 5a. The disappearance of zonal components of acceleration at the equator was already noted in connection with Fig. 4. The ratio of zonal to meridional components rises from zero, at the equator, to approach a value of unity, at the highest latitudes. The meridional components may thus presumably exert greater influence on circulations in the tropical regions, whereas the zonal and meridional components may contribute more or less equally at latitudes poleward of 60° north and south.

Figure 5b illustrates the variability of the acceleration magnitude as a function of latitude. Means of the absolute values of the accelerations for each latitude circle are compared with the peak acceleration for the field as a whole. The relatively lower mean magnitude for the tropical latitudes may be attributed to the fact that points located on the equator experience zero acceleration conditions approximately twice per day, which is not the case for locations situated away from the equator.

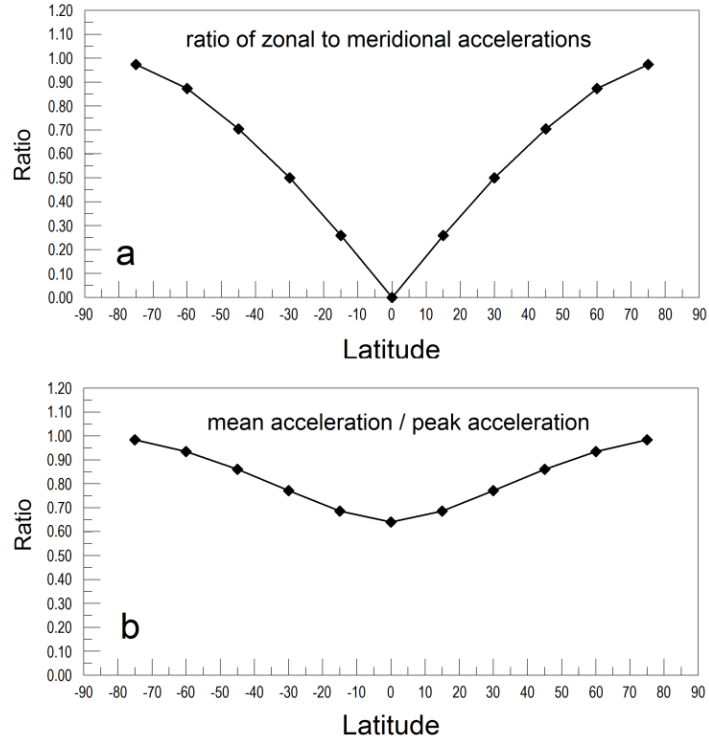

**Figure 5**. Variability of zonal and meridional acceleration components and magnitudes as a function of latitude. Data for these figures was obtained using the equally spaced, globally distributed grid points displayed in Fig. 4. A: Ratio of the magnitudes of the east-west (zonal) components of acceleration to the north-south (meridional) components. The zonal components disappear at the equator of the subject body (Fig. 4), and are systematically smaller than the meridional components everywhere except at high latitudes. B. Averaged acceleration vector resultant magnitudes are ratioed with the peak acceleration value for the field as a whole.

We conclude from Fig. 5 that the potentially observable effects and consequences of the acceleration field of Fig. 4 may differ significantly as a function of latitude. Two important factors are noted: 1) The meridional components of acceleration are twice as large as the zonal components, when averaged globally; and 2) the dominance of the meridional components over the zonal components is expressed most strongly in the equatorial and tropical latitudes.

*4.2. Temporal variability of the forcing function ($\dot{L}$)*

The orbital angular momentum of Mars, and its time derivative, as illustrated in Fig. 6, may be determined with great accuracy for periods of many hundreds of years before and after the present time [*Shirley and Mischna*, 2016]. The $\dot{L}$ waveform, shown in blue in Fig. 6, varies quasi-sinusoidally, with a mean period of ~2.2 Earth yr [*Shirley*, 2015]. This imparts a *seasonal time scale variability* to the acceleration field that is superimposed on the approximately diurnal variability described in connection with Fig. 4 above (Section 3.3). That is, in addition to the cyclic "quasi-diurnal" variability of the acceleration magnitudes and directions at a specific location that is linked with the planetary rotation, we must also envision the acceleration field pulsating in magnitude, and periodically disappearing entirely, when the forcing function ($\dot{L}$) attains zero values. The field subsequently re-emerges, with reversed sign, following the zero-crossing times of the forcing function, as illustrated in Fig. 6.

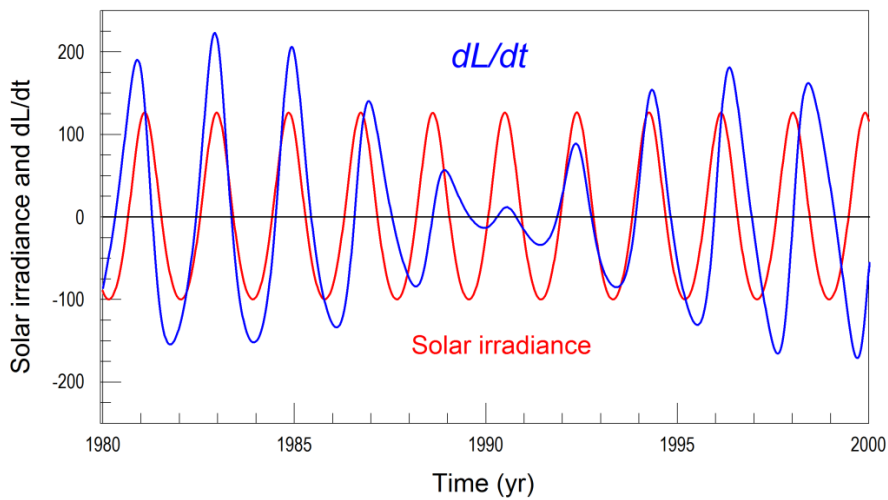

**Figure 6**. Phasing of the waveforms of the rate of change of the orbital angular momentum of Mars (*dL/dt*, in blue) and the solar irradiance at Mars (in red) from 1980-2000. Units on the y-

axis correspond to deviations from the mean value of the solar irradiance at Mars (~576 W m$^2$). The *dL/dt* (i.e., $\dot{L}$) waveform has been arbitrarily scaled for purposes of illustration. It attains positive and negative extrema of 4.45 x10$^6$ M$_{Mars}$ m$^2$ s$^{-2}$ and -3.41 x10$^6$ M$_{Mars}$ m$^2$ s$^{-2}$ during this interval. The variability of the amplitude and period of the $\dot{L}$ waveform is due to solar system dynamics, as described in *Shirley* [2015]. The phasing of the $\dot{L}$ waveform with respect to the standard solar Mars year (in red) is discussed more extensively in *Shirley and Mischna* [2016; see also *Mischna and Shirley*, 2016].

The seasonal-time-scale temporal variability of the $\dot{L}$ waveform plays a crucial role in the discussion of hypothesis testing and predictability that follows in Section 5. In particular, the diminution and disappearance of the acceleration field near the zero-crossing times is an outcome that may be evaluated through numerical modeling. Similarly, the reversal of the *sign* of the couple (with each of the acceleration vectors of Fig. 4 reversing direction) is likely to give rise to detectable differences of system response, if all other factors are held equal. Intervals in which the $\dot{L}$ waveform is positive correspond to periods when Mars is gaining orbital angular momentum, at the expense of other members of the solar system family; a substantially different atmospheric system response may occur during intervals when Mars is losing orbital angular momentum. For ease of discussion, we will later (and in companion papers) refer to these diametrically opposed conditions as *positive and negative polarity episodes*, respectively.

*4.3. Acceleration magnitude and the coupling efficiency coefficient*

The acceleration magnitudes (Fig. 4) are additionally scaled by the adopted value of the coupling coefficient *c*. As noted in Section 3.5, we consider this to represent a fundamental property of individual subject bodies which may be determined empirically. The choice of *c* is presently best constrained through comparisons of numerical modeling outcomes with observations; if it is too large, an unrealistic system response will occur. Conversely, little or no system response will be noted if the adopted value is too small. (The use of nonzero values of *c* may only be justified when an improved correspondence of model outcomes with observations is thereby achieved).

A preliminary estimate for the value of *c* for the case of the Mars atmosphere is obtained in a companion paper [*Mischna and Shirley*, 2016]. A value of $c=5.0 \times 10^{-13}$ was adopted for that investigation. The resulting peak value of the acceleration of the Mars atmosphere found for any locality over the interval 1920-2030 was ~ $2.23 \times 10^{-4}$ m s$^{-2}$. Numerical modeling [*Mischna and Shirley*, 2016] demonstrates that this small acceleration, acting on seasonal timescales, is sufficient to give rise to significant differences of near-surface Mars wind speeds. The quantitative comparisons of the following section help to place these numbers in proper perspective.

5. **What level of angular momentum exchange is permitted by solar system observations?**

Ephemerides developed for use in planetary exploration attain remarkable levels of accuracy and precision [*Giorgini et al.*, 1996; *Georgini and Yeomans*, 1999; *Folkner et al.*, 2009, 2014; *Giorgini*, 2015]. Formal statistical uncertainties in the trajectories of Earth and Venus during the current era are at the meter level; that of Mars has somewhat greater uncertainties, due in part to the influence of main-belt asteroids, whose masses are imperfectly known. Aside from the effects of dissipative tidal friction on the lunar motion, no account is taken of any possible exchange between orbital and rotational reservoirs of angular momentum in the production of the ephemerides. Thus the success of the ephemerides provides important constraints for orbit-spin coupling hypotheses, as the uncertainties of the solutions for positions and velocities can help us identify an upper bound for the permissible levels of angular momentum exchange, and thereby for the efficiency of the coupling (*cf.* Sections 2.3, 3.6, 4.3). It is thus important to understand the levels of uncertainty of the orbital angular momenta that are implied by our best solutions for planetary motions.

Recalling once more that the angular momentum may be obtained from $\boldsymbol{L} = \mathrm{M}\boldsymbol{V}\boldsymbol{R}$, where M is the mass of the subject body, $\boldsymbol{V}$ is the orbital velocity, and $\boldsymbol{R}$ is the distance separating the center of the subject body from the system barycenter, we immediately ask: At what level is $\boldsymbol{L}$ affected by meter-level errors in position knowledge $\boldsymbol{R}$? Mars orbits at a mean distance of ~$2.27939 \times 10^{11}$ m from the Sun. An error of 1 m represents a fractional proportion of $4.4 \times 10^{-12}$ of this quantity. Clearly if an exchange of angular momentum was occurring between the orbital and rotational reservoirs at this fractional level or below, we would not be able to detect its presence through presently available astronomical solutions, as the consequent variability of the orbital motions would reside "in the noise." (From the standpoint of a formal error analysis, even larger levels of exchange might be considered possible, on the basis of our knowledge of the mass of Mars, which is known only to about 5 decimal places, at ~ $6.41712 \times 10^{23}$ kg [*Folkner et al.*, 2014]).

We conclude on the basis of this calculation that the maximum permissible level of angular momentum exchange between the orbital and rotational reservoirs of Mars is likely to be no greater than about 1 part in $10^{-12}$ of $\boldsymbol{L_{Mars}}$, or ~ $3.5 \times 10^{27}$ kg m$^2$ s$^{-1}$. Quantities of angular momentum far smaller than this may be geophysically consequential. For perspective, the *total* angular momentum of Mars' atmosphere considered separately is on the order of $10^{24}$ kg m$^2$ s$^{-1}$ [*Karatekin et al.*, 2011]. Clearly there is considerable scope and leeway for significant geophysical consequences and effects, at angular momentum exchange levels well below those that are permissible on the basis of solar system observations, given the precision of the best available ephemerides.

**6. Predictability**

With reference to the terrestrial planets with atmospheres (Venus, Earth, and Mars), we suspect that the acceleration field arising due to the coupling term accelerations will not typically represent the dominant forcing function for atmospheric motions, but will instead more often comprise a perturbation superimposed on circulation patterns that are established and maintained by the seasonal variability of the solar forcing. While the structure of the acceleration field (Fig. 4) is relatively simple, the consequences of the introduction of such a field within the context of the already complex dynamics of an extended body atmosphere may be nonlinear and counter-intuitive. Even so, from the discussion of Section 4, it is possible to draw certain inferences concerning the likely behaviors of atmospheric systems perturbed by the coupling term accelerations (CTA). We here wish to identify system behaviors that may be employed in subsequent investigations for validation or disqualification of the present hypothesis.

*6.1. Constructive and destructive interference effects*

Generally speaking, if $c \neq 0$ and the accelerations are of an appreciable magnitude, the CTA will (nearly) everywhere locally modify the pre-existing velocities of the particles or parcels of atmosphere within the atmospheric system under investigation. The accelerations constructively or destructively interfere with the atmospheric state of motion that would otherwise prevail in the absence of the CTA. This is the most fundamental and most direct predictive statement that can be made with confidence for all possible situations. This outcome may be difficult to isolate and quantify through direct observations, however. Observable consequences may take on a number of forms, as discussed below.

*6.2. Range of variability*

Considering the case of two otherwise identical model atmospheres, one with and one without the CTA, we may envision the following consequence that may be of some utility in assessing the outcomes of numerical modeling experiments. The range of variability of conditions produced in current numerical models is often observed to be less than that found in the corresponding real atmosphere [*Forget et al.*, 1999; *Richardson et al.*, 2007]. In particular, the seasonal cycles of numerical models tend to show greater repeatability than is typically observed in nature. By introducing a new source of acceleration within a model atmosphere, with variable phasing with respect to seasonal cycles of solar irradiance (Fig. 6), we may expect to encounter an increased range of variability of model outcomes. Such variability may be expressed in greater variability about the mean values of globally averaged quantities, or in greater or lesser degrees of meridionality, or storminess and baroclinicity, or in other indices.

*6.3. Intensification and relaxation of circulatory flows*

In a comparison between otherwise identical atmospheres, one with and one without the accelerations, it seems likely that the accelerated atmosphere will exhibit cycles of overall intensification and relaxation of circulatory flows (presuming that $c \neq 0$ and that the accelerations are of an appreciable magnitude in comparison with other forces acting). Our hypothesis implies that the peaks of the intensification phases for the subject atmosphere should correspond in time to the extrema of the $\dot{L}$ waveform of Fig. 6, while the relaxation phases will correspond to the times of disappearance of all coupling term accelerations, when $\dot{L}$ is near zero values.

An important caveat is that we must avoid a too-literal interpretation of the term 'intensification.' The accelerations are not uniform everywhere, but instead exhibit substantial variability as a function of latitude, longitude, and time. While it is possible that a strictly proportional acceleration of some pre-existing large-scale circulatory flow (such as a western boundary current) might occasionally occur in response to the CTA, it is more likely that the adjustment of the atmosphere will take the form of structural or morphological changes, including modified patterns of large scale flows. *Destructive interference* with regular seasonal atmospheric flow patterns may occur under some circulatory 'intensification' conditions. Global indices, such as the velocities of zonal or meridional flows taken separately, may not necessarily exhibit cycles of intensification and relaxation that linearly track the waveform of the driving function ($\dot{L}$; Fig. 6). Thus while cycles of circulatory intensification and relaxation, with phasing linked with the $\dot{L}$ waveform, may be an observable outcome in many cases, we clearly cannot reasonably assert that this will occur in all possible situations, for all measurable indices, without exception.

Despite these caveats, the introduction of cycles of intensification and relaxation of circulatory flows within a model atmosphere represents an outcome that is relatively unambiguous and is measurable, both through comparisons of forced model runs with unforced model outcomes, and through comparisons of model results with atmospheric observations [*Mischna and Shirley*, 2016].

*6.4. Criteria for success or failure*

The foregoing discussion of observable consequences and predictive statements frequently mentions comparisons between numerical model runs performed with and without the coupling term accelerations. The results of such comparisons would be of little more than academic interest, if no measurable increment in the fidelity of our model results with respect to observations was thereby gained. As hinted earlier, our principal criterion for success in testing goes somewhat beyond this:

*The fundamental metric for the success or failure of the orbit-spin coupling hypothesis is whether numerical models incorporating the coupling term accelerations may thereby achieve improved correspondence of model results with observations.*

A key goal during initial hypothesis testing will thus be to determine whether a nonzero value of $c$ exists that may materially improve the correspondence of model outcomes with observations. Further, if the orbit-spin coupling hypothesis is substantially correct, then we may expect that the value of $c$, when optimized, will help constrain quantitative aspects of the underlying processes that mediate the suspected ongoing transfer of momentum between the orbital and rotational reservoirs.

## 7. Discussion

The objectives of this paper have been to document the method by which the orbit-spin coupling acceleration was obtained, and to describe the predictions of the orbit-spin coupling hypothesis as it relates to atmospheric circulations. Both of these objectives have now been achieved.

*7.1. First-order implications*

### 7.1.1. Implications for other geophysical sub-systems of extended bodies

While our presentation has focused exclusively on *atmospheric* phenomena, the physical hypothesis may carry significant implications for other geophysical systems and disciplines. A coupling of the orbital and rotational angular momenta of extended bodies is likely to involve multiple simultaneous responses and interactions of the co-rotating physical systems of which they are comprised; atmosphere, oceans, crust, mantle and core may each react or participate in different ways. If the hypothesis presented here is substantially correct, then these responses and interactions are likely to provide the subject matter for many future investigations.

### 7.1.2. Implications for solar physics and astrophysics

If the hypothesis presented here is substantially correct, then it is natural to suppose that orbit-spin coupling may likewise be operative in physical situations that differ significantly from the case of Mars and its tenuous atmosphere; astrophysical examples, and solar-physical applications [*cf. Shirley*, 2006] come immediately to mind.

### 7.1.3. A key implication for geophysical modeling of rotational processes and effects

As noted in Section 2.3, and discussed more fully in the Appendices, we have traditionally considered the orbital and rotational angular momenta of extended bodies to be independent and uncoupled, aside from certain small tidal effects. This "closed system" approach allows us to assume that observed changes to the rotation of the solid portions of a terrestrial planet must necessarily be compensated by equal and opposed changes in the momenta of the fluid components, thereby conserving rotational angular momentum. If the hypothesis presented here is substantially correct, then we must be prepared to abandon the closed-system assumption.

*7.2. Atmospheric kinetic energy and frictional processes*

The envisioned continuous exchange of angular momentum between the orbital and rotational reservoirs presumably contributes kinetic energy (during positive and negative polarity episodes; Fig. 6) to the atmosphere of the subject body. Kinetic energy within the atmosphere of Mars is mainly generated by differential solar heating; it is thereafter dissipated via frictional processes [*cf. Tabataba-Vakili et al*., 2015]. By contributing to the variability of atmospheric kinetic energy, we suspect that the present mechanism may likewise contribute to the variability with time of dissipative frictional energy losses from the system. At present the solar radiative input to the system is considered to represent the sole source of input energy subject to dissipation. If some portion of the energy dissipated is drawn either from the planetary rotation, or from the orbital motion, then it is possible that either or both of these reservoirs of energy may be subject to long term variability that is not presently captured by our dynamical models. This may conceivably represent an important dynamical consequence of the physical mechanism presented here.

*7.3. Additional physical implications of the coupling coefficient c*

There are many situations in universe characterized by orbital interactions that are more extreme than those with which we are familiar in the solar system. If the hypothesis presented here is substantially correct, then the possible role of dissipative interactions due to orbit-spin coupling in the long-term evolution of the orbital motions of stellar objects, and the consequent motions and evolution of large-scale structures over the lifetime of the universe, may be topics of some interest for subsequent investigation. A key physical question thus emerges, which may be stated as follows: *Does nature prefer and widely employ an orbit-spin coupling physics with nonzero c?*

*7.4. Present and future work: Mars and the Earth*

If the orbit-spin coupling hypothesis is substantially correct, then presumably the atmospheric circulations of solar system bodies other than Mars should likewise respond, either actively or passively, as angular momentum is exchanged between the orbital and rotational reservoirs of the body in question. The case of Venus, with its massive atmosphere and retrograde rotation, appears to provide a promising testbed for further modeling and experimentation. However, detailed observations of the circulation of the Venus atmosphere extending over many years are not yet available. The same is true for a number of other solar system targets, including Titan and the giant planets. Thus here we will restrict our attention to the more favorable cases of Mars and the Earth.

7.4.1. The Mars atmosphere

We have been and continue to be motivated to understand the origins of the interannual variability of weather and climate on Mars, as represented in the occurrence of global-scale dust storms in some Mars years but not in others. Substantial progress has been made in this area [*Mischna and Shirley*, 2016; this issue]. Convergent forecasts for the occurrence of a global-scale dust storm on Mars in the current Mars year, based on solar system dynamics, are presented in *Shirley* [2015], in *Shirley and Mischna* [2016], and *Mischna and Shirley* [2016]. It will be damaging, but not necessarily fatal, to the present orbit-spin coupling hypothesis, if no such storm occurs during the current Mars year (with perihelion in late 2016) or in the following Mars year (with perihelion in 2018). The predictions of the orbit-spin hypothesis, as detailed in Section 6, are now being subjected to more exhaustive and more sophisticated modeling efforts. These may better illuminate the underlying physics. The investigations performed to date have barely scratched the surface.

### 7.4.2. Earth's atmospheric circulation

Introduction of the coupling term accelerations within existing global circulation models (and within coupled ocean-atmosphere models) of the terrestrial atmosphere is a worthwhile objective. The resulting models may then be employed, after appropriate sanity checking and calibration, to explore the applicability of the orbit-spin coupling hypothesis for the special case of the Earth.

The terrestrial atmospheric circulation is considerably more complex than that of Mars, due largely to the presence of surface water in multiple forms. Feedbacks on atmospheric temperatures and winds due to seasonally varying albedo variations of clouds and the cryosphere are important. Likewise important is the memory effect of the thermal inertia of the oceans. While the radiative relaxation time ($\tau$) for the Mars atmosphere is 1-2 days [*Leovy*, 2001], that for the terrestrial atmosphere is approximately one month [*Wells*, 2007]; further, the inclusion of oceanic effects leads to a still longer system memory time constant estimate of the order of 5 years [*Schwartz*, 2007]. Superimposition of the coupling term accelerations upon a system of such complexity may lead to non-linear effects and consequences that are largely unforeseeable, at least without the benefit of numerical modeling. The "signal" due to the small added accelerations may be smeared out and otherwise disguised. The system complexity and the presence of system memory makes the task of validating the orbit-spin coupling hypothesis through numerical modeling of the terrestrial atmosphere considerably more difficult than is the case for the Mars atmosphere.

Nonetheless, while the effort required may be substantial, the potential benefits to society of an improved capability to forecast terrestrial weather and climate variability on seasonal and longer timescales would unquestionably be significant.

## 8. Conclusions

This paper introduces a deterministic physical mechanism capable of modifying and modulating large-scale flows of geophysical fluids in the atmosphere of Mars. Our derivation yields a previously unsuspected acceleration that varies with position and with time on and within extended bodies that are members of two-body or multi-body systems. The coupling term acceleration, given by $-c\,(\dot{\mathbf{L}} \times \boldsymbol{\omega}_\alpha) \times \mathbf{r}$, identifies and quantifies a possible physical connection between two macroscopic physical entities (the rate of change of orbital angular momentum, $\dot{\mathbf{L}}$, and the circulatory flows of planetary atmospheres) that have not previously been linked in any way. The global properties of the coupling term acceleration field have been detailed, and testable predictive statements have been formulated; our hypothesis predicts cycles of intensification and relaxation of circulatory flows of atmospheres on seasonal and longer timescales that are largely independent of solar forcing. We have shown that the operation of the mechanism envisioned does not violate constraints deriving from astronomical observations or from existing high-precision calculations of planetary motions. Proof of concept has now been obtained through numerical modeling, employing the MarsWRF GCM [*Mischna and Shirley*, 2016]. Further development and testing of this physical hypothesis may lead to an improved understanding of the fundamental causes of interannual variability of weather and climate on Mars and on the Earth.

**Appendix 1: Independence of orbital and rotational motions**

*"Kinematically, the motion of the Earth as a whole can be represented as the resultant of a translation and a rotation in an indefinite number of ways. The particular one that is most advantageous from a dynamical point of view is a representation as the resultant of a translation of the Earth as a whole with the velocity of the center of mass, and a rotation about an axis through the center of mass. These two component motions are **dynamically independent of each other**…"* E. W. Woolard [1953; *emphasis added*].

Separate consideration of the orbital motions and rotational motions of stars and planets has served dynamical astronomy well. From a practical standpoint, treating these two forms of motion separately offers a great simplification for calculations. Isaac Newton found that for most dynamical purposes, extended bodies could be treated as point-masses [*Newton*, 1687]. This removed any requirement to additionally consider the form and rotation state of extended bodies interacting gravitationally at great distances.

In Newtonian theory, and in relativistic theory in the Newtonian limit, conservation of angular momentum is considered to hold separately for orbital angular momentum and rotational angular momentum. However, when estimating the total angular momentum of the solar system, it is usual to sum the values from each category to produce a single value.

In metric theories, the question of the separability and independence of orbital and rotational motions is largely obviated by the representation of orbital motions as a state of free fall along geodesic trajectories of curved spacetime [*Misner, Thorne and Wheeler*, 1973; *Shirley*, 2006], in which no force is associated with the orbital motion.

**Appendix 2: Gravitation and rotation**

The rotational motions of extended bodies may be influenced by gravitational phenomena in a number of ways. Some of the interactions noted here, such as the luni-solar precession of the Earth, are conservative, without long-term evolutionary consequences for orbits. The effects and consequences of tidal friction, which is dissipative, are also briefly summarized in this Appendix.

*A2.1. Conservative coupling mechanisms*

Gravitational torques acting on the equatorial bulges of rotating bodies give rise to the linked phenomena of precession and nutation, which were known to astronomers in antiquity. For aspherical satellites in synchronous rotation, such as Earth's Moon, the same torques excite small periodic oscillatory motions known as physical librations. Precession-nutation and the physical librations are considered to be conservative, having no long-term evolutionary consequences for the orbit or the rotation rate of the affected body. Precession and nutation are fundamentally due to tidal gradients of external gravitational potentials experienced by a subject body [*Melchior*, 1971; *Lambeck*, 1988; see also Appendix 3].

*A2.2. Tidal friction*

Tidal friction is an important dissipative mechanism that links gravitational accelerations arising in the dynamical environment with secular changes in the rotation states of extended bodies. Here a torque arises due to gravitational forces acting on lagged (or misaligned) tidal bulges of rotating bodies [*Munk and MacDonald*, 1960; *Hipkin*, 1970; *Lambeck*, 1980; *Wahr*, 1985]. Elastic forces account for displacements of the tidal bulges away from the line connecting the centers of the interacting bodies. Tidal friction is invoked to explain the linked phenomena of an observed slow secular expansion of the lunar orbit, together with an observed long-term slowing of the rate of rotation of the Earth [*cf. Counselman*, 1973; *Calame and Mulholland*, 1978; *Hansen*, 1982; *Brosche*, 1984]. This mechanism is dissipative, in that some rotational energy is lost due to frictional processes, principally within the Earth's oceans [*Lambeck*, 1980, 1988, and references therein].

An extensive literature explores the applicability and effects of tidal friction and associated phenomena in the Earth [*Lambeck*, 1980, 1988, and references therein], in the planets and satellites of the solar system [*Goldreich and Peale*, 1966; *Peale*, 1999, and references therein], and in some stars [*Zahn*, 1977; *Eggleton et al.*, 1998] and extra-solar planet systems [*Rodriguez et al.*, 2012, and references therein]. The synchronous rotations of many of the large satellites of the solar system, including Earth's Moon and the Galilean satellites of Jupiter, are considered to represent an end result of this form of coupling of orbital and rotational motions [*Goldreich and Peale*, 1966; *Burns*, 1986; *Peale*, 1999]. In extreme cases, significant geophysical effects are noted; among these is the flexural tidal heating of Jupiter's satellite Io, which contributes to the generation of the continuous volcanic activity of that body [*Peale et al.*, 1979; *Greenberg*, 1989; *Moore*, 2003].

The torques associated with tidal friction show an inverse sixth power dependence on distance [*cf. Hipkin*, 1970], and so, at present, this form of coupling is largely relevant only for relatively nearby satellites of the Sun and planets [*Wilkins and Sinclair*, 1974], and for close companions in stellar systems [*Rodriguez et al.*, 2012, and references therein].

*A2.3. Summary:*

Aside from the above effects resulting from tides, we presently know of no other process arising within the solar system dynamical environment that may actively couple the orbital and rotational motions of extended bodies.

**Appendix 3: Gravitational tides on Mars**

Gravitational tides are proportional to the inverse third power of distance, and so fall off rapidly on the distance scales appropriate for solar system investigations. The vertical component of tidal acceleration for a point located on the surface of an extended body immediately beneath the disturbing body may be obtained from

$$a_{tidal} = (2GM m \mathbf{r})/\mathbf{R}^3 \qquad (A3.1)$$

where $G$ is the constant of gravitation ($6.673 \times 10^{-11}$ m$^3$ kg$^{-1}$ s$^{-2}$), M is the mass of the disturbing body, m is an element of mass of the disturbed body, $\mathbf{r}$ is the radius of the subject body, and $\mathbf{R}$ is the distance separating the centers of the two. Making use of planetary mass values and mean orbital radii from *Seidelmann* [1992], we can obtain and compare representative tidal accelerations for some of the most important tide-raising bodies for Mars, as identified in Table A3-1.

| Gravitational Tides on Mars | |
|---|---|
| Source | Tidal acceleration (m sec-2) |
| Sun | 7.62 e-08 |
| Phobos | 5.29 e-09 |
| Deimos | 6.76 e-11 |
| Earth | 5.63 e-12 |
| Jupiter | 5.16 e-12 |
| Saturn | 1.49 e-13 |
| Ceres | 6.61 e-17 |
| Vesta | 5.96 e-17 |

**Table A3-1**. Vertical component of the gravitational tidal accelerations on Mars due to the Sun, the moons of Mars, the Earth, Jupiter, Saturn, and two large asteroids, as obtained using equation (A2.1), for a surface location directly beneath the tide-generating body.

The Sun raises by far the largest gravitational tide on Mars, followed by Phobos (smaller by a factor of 14), and Deimos (smaller by a factor of 1127). We can place these accelerations in perspective through a comparison with those obtained in a similar manner for the Earth. When this is done, we find that the largest tidal acceleration on Mars is more than an order of magnitude smaller than the largest (lunar) tide on Earth (i.e., by a factor of ~14).

It is widely acknowledged that the gravitational tidal accelerations on Earth are too small to substantially impact Earth's atmospheric circulation and thereby Earth's weather and climate. Tidal equations are not normally included in atmospheric global circulation models, either for the Earth or for Mars. These quantitative comparisons appear to rule out a significant role for gravitational tides in modulating the circulation of the Martian atmosphere.

a "reanalysis" of spacecraft observations, *Geophys. Res. Lett*. 42, 8320-8327, 10.1002/2015GL065659.

Wang, H., M. I. Richardson, R. J. Wilson, A. P. Ingersoll, a. D. Toigo, and R. W. Zurek, 2003. Cyclones, tides, and the origin of a cross-equatorial dust storm on Mars, *Geophys. Res. Let*t. 30, 10.1029/2002GL016828.

Wang, H. and M. I. Richardson, 2015. The origin, evolution, and trajectory of large dust storms on Mars during Mars years 24-30 (1999-2011), *Icarus* 251, 112-127, doi:10.1016/j.icarus.2013.10.033.

Wells, N.C., 2007. The atmosphere and ocean: A physical introduction, §1.3-1.4, Wiley, NY (448 pps).

Wilkins, G. A., and A. T. Sinclair, 1974. The dynamics of planets and their satellites, *Proc. Roy. Soc. Lond. A*, 336, 85-104.

Wilson, R. J., 2002. Evidence for nonmigrating thermal tides in the Mars upper atmosphere from the Mars Global Surveyor Accelerometer Experiment, *Geophys. Res. Lett*. 29, 10.1029/2001GL013975.

Wilson, R. J., and K. P. Hamilton, 1996. Comprehensive model simulation of thermal tides in the Martian atmosphere, *J. Atmos. Sci*. 53, 1290-1326.

Withers, P., R. Pratt, J.-L. Bertaux, and F. Montmessin, 2011. Observations of thermal tides in the middle atmosphere of Mars by the SPICAM instrument, *J. Geophys. Res*. 116, E11005, 10.1029/2011JE003847.

Woolard, E. W., 1953. *Theory of the rotation of the Earth around its center of mass*, Astron. Pap. Am. Eph. 15, Part 1, pp. 11-162.
57

**10. Acknowledgements**: Development of the ideas and approach described in this manuscript occurred over a time span of many years. I thank H. T. Mörth, R. W. Fairbridge, G. H. Dunne, and D. J. McCleese for criticism and for encouragement over many years. I thank M. A. Mischna, Jon Giorgini, Eric Rosenberg, Nick Heavens, and Claire Newman for suggestions and for critical reviews of earlier versions of this paper. Discussions with I. Charvátová, D. Juckett, R. W. Zurek, Tim Schofield, and the members of the Mars Climate Sounder Science Team have materially improved this presentation. Portions of this work were performed at the Jet Propulsion Laboratory, California Institute of Technology, under a grant from NASA. Support from NASA's Solar System Working Program (Grant 14-SSW14_2-0354) is gratefully acknowledged.